\documentclass[a4paper,11pt]{article}

\usepackage{amsmath}
\usepackage{amssymb}
\usepackage[dvips]{graphicx}

\makeatletter
\@addtoreset{equation}{section}
\renewcommand{\theequation}{\thesection.\@arabic\c@equation}
\makeatother

\makeatletter
\renewcommand\appendix{\par
  \setcounter{section}{0}%
  \setcounter{subsection}{0}%
  \gdef\thesection{Appendix \@Alph\c@section }
  \renewcommand{\theequation}
  {\Alph{section}.\arabic{equation}}
}
\makeatother

\def \be {\begin{equation}}
\def \ee {\end{equation}}
\def \ba {\begin{array}}
\def \ea {\end{array}}
\def \bea{\begin{eqnarray}}
\def \eea{\end{eqnarray}}

\def \sch {Schwarzschild~}

\begin{document}
\begin{titlepage}
\vspace{1cm} 
\begin{center}
{\Large \bf {Higher Spins Tunneling from a Time Dependent and Spherically Symmetric Black Hole}}\\
\end{center}
\vspace{2cm}
\begin{center}
\renewcommand{\thefootnote}{\fnsymbol{footnote}}
Haryanto M. Siahaan{\footnote{haryanto.siahaan@unpar.ac.id}}
\\
Physics Department, Parahyangan Catholic University,\\
Jalan Ciumbuleuit 94, Bandung 40141, Indonesia

\renewcommand{\thefootnote}{\arabic{footnote}}
\end{center}
%
\begin{abstract}
The discussions of Hawking radiation via tunneling method have been performed extensively in the case of scalar particles. Moreover, there are also several works in discussing the tunneling method for Hawking radiation by using higher spins, e.g. neutrino, photon, and gravitino, in the background of static black holes. Interestingly, it is found that the Hawking temperature for static black holes using the higher spins particles has no difference compared to the one computed using scalars. In this paper, we study the Hawking radiation for a spherically symmetric and time dependent black holes using the tunneling of Dirac particles, photon, and gravitino. We find that the obtained Hawking temperature is similar to the one derived in the tunneling method by using scalars. 
%
%
\end{abstract}
\end{titlepage}\onecolumn 
\bigskip 
\section{Introduction}
General relativity predicts the existence of black holes, i.e. objects from which nothing can escape. Hawking found that incorporating quantum mechanics changes this understanding about black holes \cite{Hawking1,Hawking2}. He showed that black holes are not completely black. They radiate thermal spectrum due to the quantum effects. Furthermore, Hawking found a connection between the surface gravity of a black hole and its corresponding temperature. 

There are several ways in understanding the mechanism of black holes to radiate. One of the latest approaches was given by Parikh and Wilczek\cite{Wilczek,Parikh}, where the radiation of black holes is described  as a quantum tunneling effect of particles through the event horizon. This method is found to be simpler and more intuitive. In the Parikh-Wilczek (PW) method, which is called sometime as the radial null geodesic method, one first computes the across horizon tunneling amplitude as an exponentiation of the imaginary part of the corresponding particle's action in the outgoing mode. Then the principle of detailed balance is used to connect the tunneling amplitude with the Boltzmann factor, hence the temperature can be obtained. 

In fact, the radial null geodesic method is not the only way to describe the Hawking radiation by using the tunneling mechanism. There is an alternative, known as the Hamilton-Jacobi method \cite{Paddy}, where one solves the semiclassical equation of motion for the tunneled particles by using the Hamilton-Jacobi ansatz for the corresponding particle's wave function. The discussions for Hamilton-Jacobi method in discussing the Hawking temperature and black hole thermodynamics for various cases and black holes have been performed in \cite{Vanzo,Kim,Jiang,Banerjee-Majhi,Bibhas,RB}. However, most of them are confined into the tunneling of scalar fields where one starts with a Klein-Gordon equation in curved spacetime and solves it by using Hamilton-Jacobi ansatz. 

For static spacetimes, the tunneling mechanism for Hawking radiation has been extended to the case of spin $\tfrac{1}{2}$ fermion \cite{Mann-spin}, as well as photon and gravitino \cite{Majhispin}. These works motivate us to study the Hawking mechanism for time dependent black hole's radiation as the tunneling processes of massless higher spins particles. The works presented in this paper can be also considered as an extension of our previous work \cite{Siahaan:2009qv} where the authors study the scalar tunneling which gives rise to the Hawking temperature of Vaidya black holes. Since we keep the spherically symmetric and time dependent spacetime metric in this paper to be quite general, i.e. spacetime whose line element can be written as eq. (\ref{metric-gen-timedep}), hence the results presented in this paper should be relevant to this class of spacetime\footnote{For example to those discussed in \cite{dynamicalBH}.}.

The organization of paper is as follows. In the next section, we review the null geodesic method for a general time dependent metric background. In section \ref{sec:Dirac}, we discuss the tunneling of massless Dirac particles across the time dependent black hole's horizon. By using the solutions of massless Dirac fields and then the detailed balance principle, we can get the Hawking temperature of the black holes under consideration. Interestingly, the Hawking temperature derived in this section is invariant compared to the one obtained in the case of tunneled scalar fields \cite{Siahaan:2009qv}. The same prescription is repeated in section \ref{sec:photon}, where the starting point is the source free Maxwell equation in curved space. Again, the Hamilton-Jacobi ansatz is used to solve the corresponding equation. After using the Lorentz gauge condition, we can get the solution for the vector fields. The same finding is appeared in section \ref{sec:photon}, where the obtained Hawking temperature has no difference to that derived in the scalar tunneling \cite{Siahaan:2009qv}. 

In section \ref{sec:gravitino}, the tunneled object is massless gravitino, i.e. particle with spin $\tfrac{3}{2}$. We start from the Rarita-Schwinger equation in curved space, and Hamilton-Jacobi ansatz helps us to get the solutions. As we would expect after performing the analysis in sections \ref{sec:Dirac} and \ref{sec:photon}, the Hawking radiation due to the massless gravitino tunneling yields the same temperature as one in the scalar case \cite{Siahaan:2009qv}. The conclusions are given in the last section.

\section{Radial Null Geodesic Method}\label{sec:radial-null}

In \cite{Wilczek}, Parikh and Wilczek presented a direct and short derivation of Hawking radiation as a tunneling process. To dispel the coordinate singularity that the \sch coordinate has, they employ the Painleve transformation, which yields the \sch spacetime transforms to
\be
ds^2  =  - \left( {1 - \frac{{2M}}{r}} \right)dt^2  + 2\sqrt {\frac{{2M}}{r}} dtdr + dr^2  + r^2 d\Omega _2^2 \,,
\ee 
where $d\Omega_2 ^2=d\theta^2 + \sin^2\theta d\phi^2$ is the metric of 2-sphere with unit radius. In getting the last line element from the \sch metric, the \sch time $t_s$ is transformed as \cite{Wilczek}
\be\label{Painleve}
t_s  = t - 2\sqrt {2Mr}  - 2M\ln \left( {\frac{{\sqrt r  - \sqrt {2M} }}{{\sqrt r  + \sqrt {2M} }}} \right)\,.
\ee 
For a general spherically symmetric and static spacetime
\be \label{static-gen}
ds^2  =  - X\left( r \right)dt^2  + Y\left( r \right)^{ - 1} dr^2  + r^2 d\Omega _2^2 \,,
\ee 
the Painleve transformation (\ref{Painleve}) can be written as
\begin{eqnarray}
	dt \to dt - \sqrt {\frac{{1 - Y}}{{XY}}} dr\label{eq:3}\,.
\end{eqnarray}

In this paper we work out the tunneling prescription to explain the Hawking radiation for a time dependent and spherically symmetric black holes. In general, such black hole solutions can be read as \cite{Weinberg}
\begin{eqnarray}\label{metric-gen-timedep}
	ds^2  =  - X\left( {t,r} \right)dt^2  + Y\left( {t,r} \right)^{ - 1} dr^2  + Z\left( {t,r} \right) d\Omega_2 ^2\label{eq:2}\,.
\end{eqnarray}
In this paper we study a class of time dependent spacetime which has coordinate singularity at $r$ which yields $Y(t,r)=0$ and $X(t,r)=0$. In order to get rid of this coordinate singularity, we employ the ``generalized'' Painleve transformation (\ref{eq:3}) which is in a differential form can be read as
\be \label{Painleve-diff}
dt_P = \frac{{\partial t_P}}{{\partial t }}dt  + \frac{{\partial t_P}}{{\partial r}}dr\,.
\ee
For a general spacetime metric (\ref{metric-gen-timedep}), it is clear that both ${{\partial t_P } \mathord{\left/
 {\vphantom {{\partial t_P } {\partial t}}} \right.
 \kern-\nulldelimiterspace} {\partial t}}$ and ${{\partial t_P } \mathord{\left/
  {\vphantom {{\partial t_P } {\partial r}}} \right.
  \kern-\nulldelimiterspace} {\partial r}}$ would be some functions of $t$ and $r$. However, by employing the relation which are applied in static case, i.e. $dt_P  = dt - \sqrt {\frac{{1 - Y}}{{XY}}} dr$, we find that the mapping
  \be \label{dtpdt}
  \frac{{\partial t_P }}{{\partial t}} = 1\,
  \ee
	and
  \be \label{dtpdr}
  \frac{{\partial t_P }}{{\partial r}} =  - \sqrt {\frac{{1 - Y\left( {t,r} \right)}}{{X\left( {t,r} \right)Y\left( {t,r} \right)}}} 
  \ee
   remove the coordinate singularity in the general metric (\ref{metric-gen-timedep}). As in the static case, the left hand side of eq. (\ref{dtpdr}) is singular for the vanishing $X(t,r)$ or $Y(t,r)$. This is a normal consequence in such transformation to remove the coordinate singularity. 
   
   Furthermore, to get an integrable $t_P$, from the fact that ${\textstyle{{\partial ^2 t_P } \over {\partial r\partial t}}} = 0$ in eq. (\ref{dtpdt}) one realizes that the right hand side of eq. (\ref{dtpdr}) must be independent of time $t$, say\footnote{In the case of static spacetime, eq. (\ref{int.cond}) is automatically satisfied.}
   \be\label{int.cond}
   1 - Y\left( {t,r} \right) = X\left( {t,r} \right)Y\left( {t,r} \right)C\left( r \right)\,,
   \ee 
   where $C(r)$ is an arbitrary function of $r$. Hence, a set of Painleve transformation (\ref{dtpdt}) and (\ref{dtpdr}) works only for a class of time dependent spherically symmetric spacetime whose metric functions satisfy the condition\footnote{In \ref{app.Vaidya}, we show a constraint for Vaidya black hole mass function $m(t,r)$ which comes from this condition.} (\ref{int.cond}).

Consequently, after employing the transformation (\ref{dtpdt}) and (\ref{dtpdr}), the metric (\ref{metric-gen-timedep}) now can be written as
\[ds^2  =  - X(t,r)dt^2  + 2X(t,r)\sqrt{\frac{{1 - Y(t,r)}}{{X(t,r)Y(t,r)}}} dtdr  \]
\begin{eqnarray}\label{P1}
 ~~~~~~~~~~~~~~~~~~~~~~~+ dr^2	+ Z\left( {t,r} \right) d\Omega_2 ^2\label{eq:4}\,.
\end{eqnarray}
It is understood that the coordinate singularity at $X(t,r)=Y(t,r)=0$ has already been removed in (\ref{eq:4}), where instead to be singular, now the $g_{tr}$ component of the line element above is indeterminate\footnote{We consider the case where after employing the L'Hospital's rule, one can get a non singular form of $g_{tr}$ in the metric (\ref{metric-gen-timedep}). Otherwise, the method developed in this paper might not work since $dr/dt$ in (\ref{eq:5}) could be singular.}. Accordingly, the null radial geodesics from the ``Painleve transformed'' metric (\ref{P1}) can be written as
\begin{eqnarray}
	\frac{{dr}}{{dt}} = \sqrt {\frac{X(t,r)}{Y(t,r)}} \left( { \pm 1 - \sqrt {1 - Y(t,r)} } \right),\label{eq:5}
\end{eqnarray}
and $ + ( - )$ signs denote the outgoing(ingoing) geodesics. Moreover, for a practical benefit one can Taylor expand the coefficient $X$ and $Y$ at the near horizon, i.e.
\be
	\left. {X(t,r)} \right|_t  \simeq \left. {X'(t,r_h)} \right|_t \left( {r - r_h } \right) + \left. {O\left( {\left( {r - r_h } \right)^2 } \right)} \right|_t \label{eq:6}\,,
\ee
and
\be
	\left. {Y(t,r)} \right|_t  \simeq \left. {Y'(t,r_h)} \right|_t \left( {r - r_h } \right) + \left. {O\left( {\left( {r - r_h } \right)^2 } \right)} \right|_t \label{eq:7}\,,
\ee
where $r_h$ is the radius of event horizon. By using the Taylor expansions (\ref{eq:6}) and (\ref{eq:7}), the outgoing null radial geodesic (\ref{eq:5}) can be approached as
\begin{eqnarray}
	\frac{{dr}}{{dt}} \simeq \frac{1}{2}\sqrt {X'\left( {r_h ,t} \right)Y'\left( {r_h ,t} \right)} \left( {r - r_h } \right).\label{eq:8}
\end{eqnarray}

Now we use the prescription by Parikh and Wilczek in getting the Hawking temperature using the picture of tunneling particle through the event horizon, which is sometime called the Radial Null Geodesic or Parikh-Wilczek (PW) method. However, it was found by Chowdhury in \cite{Chowdhury:2006sk} that the expression ${\rm Im} S = \int p_r dr$ which appears in original PW method is not canonically invariant. Therefore, we use the prescription by Akhmedov et al \cite{Akhmedov:2008ru} in computing the tunneling rate which reads
\be\label{rate-tunn}
\Gamma  \sim \exp \left[ { - \frac{{{\mathop{\rm Im}\nolimits} \oint {p_r dr} }}{\hbar }} \right]\,.
\ee 
The term inside of square bracket above then can be computed as
\begin{eqnarray}\label{eq:9}
	 {\mathop{\rm Im}\nolimits} \oint {p_r dr}  = {\mathop{\rm Im}\nolimits} \oint {\int\limits_0^{p_r } {dp_r '} dr}  = {\mathop{\rm Im}\nolimits} \oint {\int\limits_0^H {\frac{{dH'}}{{{\textstyle{{dr} \over {dt}}}}}} dr}\,, 
\end{eqnarray}
where we have made use of the Hamilton equation $dr/dt = dH/dp_r |_r 
$ related to the canonical variables $r$ and $p_r$ (in this case, the radial component of the radius and the momentum). Different from the discussions of several authors for a static black hole mass, e.g. Refs. \cite{Wilczek} and \cite{22}, the outgoing particle's energy must be time dependent for black holes with varying mass. So, the $dH'$ integration at (\ref{eq:9}) is for all values of outgoing particle's energy, say from zero to $ + E\left( t \right)$.

By using the approximation (\ref{eq:8}), we can perform the integration (\ref{eq:9}). For $dr$ integration, we can perform a contour integration for the upper half complex plane to avoid the coordinate singularity $r_h$. The result is
\begin{eqnarray}
	{\mathop{\rm Im}\nolimits} \oint {p_r dr} = \frac{{4\pi E\left( t \right)}}{{\sqrt {X'\left( {r_h ,t} \right)Y'\left( {r_h ,t} \right)} }}.\label{eq:10}
\end{eqnarray}
Equalizing the tunneling rate (\ref{rate-tunn}) with the Boltzmann factor $\exp \left[ { - \beta E\left( t \right)} \right]$ for a system with time dependence of energy we obtain 
\begin{eqnarray}\label{Hawking-temp}
	T_H  = \frac{{\hbar \sqrt {X'\left( {r_h ,t} \right)Y'\left( {r_h ,t} \right)} }}{{4\pi }}.\label{eq:11}
\end{eqnarray}
This temperature is also derived by Nielsen and Yeom in \cite{Nielsen:2008kd} by using a slightly different approach of PW method for a general time dependent background.

At this point we understand that the temperature (\ref{Hawking-temp}) doesn't care about the spin of tunneled particles. As long as it follows the radial null geodesic, which we know the path of massless particles, then it will give the same contribution the temperature measured by a detector at infinity independent of the spin that it has. However, the spins that we are mentioning here are $0,1/2,1$ and $3/2$ only, since in the author's knowledge there is no works for Hawking radiation in the tunneling picture which uses spins $\ge 2$ as the tunneled particles. In the next sections, we will reproduce the temperature (\ref{Hawking-temp}) by considering the tunneling of Dirac fermion, photon, and gravitino from a spherically symmetric and time dependent black hole by using Hamilton-Jacobi method.

\section{Massless Dirac Particle Tunneling}\label{sec:Dirac}

In this section we will study the Hawking radiation of time dependent black holes where the tunneled particle has spin $\tfrac{1}{2}$. We start by writing an action describing massless Dirac fields in curved spacetime \cite{Nakahara}
\be\label{Dirac-action} 
S_\psi   = \int {d^4 x\sqrt { - g} \bar \Psi i\tilde \gamma ^\mu  \left( {\partial _\mu   + \frac{i}{2}g^{\gamma \nu } \Gamma _{\mu \nu }^\beta  \Sigma _{\beta \gamma } } \right)\Psi } \,.
\ee
The corresponding equation of motion can be written as
\be \label{Diraceqtn}
{\tilde{\gamma}} ^\mu  \nabla _\mu  \Psi  = 0
\ee
where
\be 
\nabla _\mu   = {\partial _\mu   + \frac{i}{2}g^{\gamma \nu } \Gamma _{\mu \nu }^\beta  \Sigma _{\beta \gamma } }\,,
\ee
and $\Sigma _{\alpha \beta }  = \frac{i}{4}\left[ {\gamma _\alpha  ,\gamma _\beta  } \right]$. We use ${\rm{diag}}(-,+,+,+)$ as the Minkowski metric tensor and the flat spacetime Dirac matrices $\gamma^\alpha$ are

\[\gamma ^0  = \left( {\begin{array}{*{20}c}
   i & 0  \\
   0 & { - i}  \\
\end{array}} \right)
\,\,,\,\,
\gamma ^1  = \left( {\begin{array}{*{20}c}
   0 & {\sigma ^3 }  \\
   {\sigma ^3 } & 0  \\
\end{array}} \right)\,,\]
\be
\gamma ^2  = \left( {\begin{array}{*{20}c}
   0 & {\sigma ^1 }  \\
   {\sigma ^1 } & 0  \\
\end{array}} \right)
\,\,,\,\,
\gamma ^3  = \left( {\begin{array}{*{20}c}
   0 & {\sigma ^2 }  \\
   {\sigma ^2 } & 0  \\
\end{array}} \right)\,.
\ee
The flat Dirac matrices $\gamma^\alpha$ and the ``curved'' ones $\tilde{\gamma}^\mu$ are related by $\tilde{\gamma}^\mu = e^\mu_\alpha \gamma^\alpha$. 

For the general time dependent metric (\ref{metric-gen-timedep}), the tetrads $e_\mu ^a$ can be expressed as
\be 
e_\mu ^a  = \left( {\begin{array}{*{20}c}
   {\sqrt X } & 0 & 0 & 0  \\
   0 & {1 \mathord{\left/
    {\vphantom {1 a}} \right.
    \kern-\nulldelimiterspace} \sqrt{Y}} & 0 & 0  \\
   0 & 0 & r & 0  \\
   0 & 0 & 0 & {r\sin \theta }  \\
\end{array}} \right)\label{tetrad}
\ee
where $g_{\mu \nu }  = e_\mu ^a e_\nu ^b \eta _{ab} $. Clearly $e^\mu_a$ is just the inverse of (\ref{tetrad}). Therefore, the Dirac matrices $\tilde \gamma ^\mu$ using the tetrads (\ref{tetrad}) can be written as
\be 
\tilde \gamma ^t  = \frac{i}{{\sqrt X }}\left( {\begin{array}{*{20}c}
   1 & 0 & 0 & 0  \\
   0 & 1 & 0 & 0  \\
   0 & 0 & { - 1} & 0  \\
   0 & 0 & 0 & { - 1}  \\
\end{array}} \right)\,,
\ee
\be
\tilde \gamma ^r  = {\sqrt Y }\left( {\begin{array}{*{20}c}
   0 & 0 & 1 & 0  \\
   0 & 0 & 0 & { - 1}  \\
   1 & 0 & 0 & 0  \\
   0 & { - 1} & 0 & 0  \\
\end{array}} \right)\,,
\ee 

\be 
\tilde \gamma ^\theta   = \frac{1}{r}\left( {\begin{array}{*{20}c}
   0 & 0 & 0 & 1  \\
   0 & 0 & 1 & 0  \\
   0 & 1 & 0 & 0  \\
   1 & 0 & 0 & 0  \\
\end{array}} \right)\,,
\ee

\be
\tilde \gamma ^\phi   = \frac{i}{{r\sin \theta }}\left( {\begin{array}{*{20}c}
   0 & 0 & 0 & { - 1}  \\
   0 & 0 & 1 & 0  \\
   0 & { - 1} & 0 & 0  \\
   1 & 0 & 0 & 0  \\
\end{array}} \right)\,.
\ee

Now we employ the Hamilton-Jacobi ansatz for the spinor wave function describing the massless Dirac particles. To simplify the computation, we perform the tunneling process analysis for spin up particle $\Psi_u$ and spin down one $\Psi_d$ separately \footnote{The reader should be familiar with the Dirac spinor as a direct sum $\Psi  = \Psi _u  \oplus \Psi _d $.}. Explicitly, the Hamilton-Jacobi ansatz for our spinors can be read as
\be \label{ansatzPsi-up}
\Psi _u  = \left( {\begin{array}{*{20}c}
   {\cal{A}}  \\
   0  \\
   {\cal{B}}  \\
   0  \\
\end{array}} \right)\exp \left( {\frac{i}{\hbar }S_u } \right)
\ee\be\label{ansatzPsi-down}
\Psi _d  = \left( {\begin{array}{*{20}c}
   0  \\
   {\cal{C}}  \\
   0  \\
   {\cal{D}}  \\
\end{array}} \right)\exp \left( {\frac{i}{\hbar }S_d } \right)\,,
\ee
where the function $S_u$ and $S_d$ are expanded in terms of $\hbar$ as
\be \label{expandS-Dirac}
S_k  = S_{0k}  + \hbar S_{1k}  + \hbar ^2 S_{2k}  + \hbar ^3 S_{3k}  + \dots 
\ee 
and $k$ is understood as the spin of Dirac particles under consideration, i.e. $k=u$ or $k=d$. The coefficients ${\cal A}$, ${\cal B}$, ${\cal C}$, and ${\cal D}$ in general are $t,r,\theta$ and $\phi$ dependent, as well as the corresponding actions for spins up and down $S_u$ and $S_d$ respectively. As we understand in the Dirac formalism, (\ref{ansatzPsi-up}) is the wave function for spin up fermion, and (\ref{ansatzPsi-down}) is for the spin down. In the next step, we work out the analysis for spin up only, since an analogous method can also be performed for spin down which produces the same Hawking temperature. The case of radial null geodesic yields the corresponding particle action does not vary with respect to the $\theta$ and $\phi$ coordinates. Therefore, by inserting the spin up wave function (\ref{ansatzPsi-up}) into the equation (\ref{Diraceqtn}), we have a set of equations
\be\label{d1}
\frac{{i{\cal{A}}}}{{\sqrt X }}\partial _t S_{0u}  + {\sqrt Y }{\cal{B}}\partial _r S_{0u}  + {\cal{O}}\left( \hbar  \right) = 0\,,
\ee

\be\label{d3}
{\sqrt Y }{\cal{A}}\partial _r S_{0u}  - \frac{{i{\cal{B}}}}{{\sqrt X }}\partial _t S_{0u}  + {\cal{O}}\left( \hbar  \right) = 0\,,
\ee
We focus only on the leading order terms in the equations, hence ${\cal{O}}(\hbar)$ can be neglected. 

Moreover, the equations (\ref{d1}) and (\ref{d3}) can be shown in a matrix operation as following
\be 
\left( {\begin{array}{*{20}c}
   {iX^{ - 1/2} \partial _t S_{0u} } & {Y^{ - 1/2} \partial _r S_{0u} }  \\
   {Y^{ - 1/2} \partial _r S_{0u} } & { - iX^{ - 1/2} \partial _t S_{0u} }  \\
\end{array}} \right)\left( {\begin{array}{*{20}c}
   {\cal{A}}  \\
   {\cal{B}}  \\
\end{array}} \right) \equiv {\tilde D}\left( {\begin{array}{*{20}c}
   {\cal{A}}  \\
   {\cal{B}}  \\
\end{array}} \right) = 0\,.
\ee
The vanishing of last equation is guaranteed if the determinant of $\tilde{D}$ is zero, which leads to
\be \label{actionSu-eq1}
\left( {\partial _t S_{0u} } \right)^2  = XY\left( {\partial _r S_{0u} } \right)^2 \,.
\ee
We notice that equation (\ref{actionSu-eq1}) is just the equation for scalar particle's action in curved space after we employ the Hamilton-Jacobi ansatz and consider only the leading terms in the equation \cite{Paddy,Siahaan:2009qv}. Moreover, the last equation can be rewritten as
\be\label{eqfn} 
\partial _r S_{0u}  =  \pm {\frac{1}{\sqrt{XY}}} \partial _t S_{0u} 
\ee
where $(-)+$ signs correspond to the (outgoing)ingoing modes. The discussions of these modes can be found in Apendix 2.A of \cite{Majhi:thesis}. 

In \cite{Siahaan:2009qv}, the authors have derived the solution for an equation like (\ref{eqfn}) where they consider the tunneling of scalar particles from a time dependent black hole. Therefore, the techniques presented in \cite{Siahaan:2009qv} can be adopted to get an expression for $S_u$. We look for a general form of solution for the action\footnote{A discussion of Schrodinger equation with time dependent Hamiltonian which supports this general form of action is given in \ref{app.timedepSchrodinger}.}
\be \label{Su-gen}
S_{0u} \left( {t,r} \right) = - \int\limits_0^t {E\left( {t'} \right)dt'}  + \tilde S_{0u} \left( {t,r} \right)\,,
\ee
where $E(t')$ stands for the time dependent energy of the Dirac particle which tunnels across the event horizon. The time dependence of energy is understood since the mass of black hole decreases as time passes. Taking the derivative with respect to time in both sides of the last equation provides us
\be 
\partial _t S_{0u} \left( {t,r} \right) = - E\left( t \right) + \partial _t \tilde S_{0u} \left( {t,r} \right)\,,
\ee
and from the differentiation with respect to radius $r$ we have
\be \label{drS-dtS}
\partial _r S_{0u} \left( {t,r} \right) = \partial _r \tilde S_{0u} \left( {t,r} \right)\,.
\ee

The chain rule allows us to write
\be \label{eqtn-Su-afterchain}
\frac{{d\tilde S_{0u} \left( {t,r} \right)}}{{dr}} = \frac{{\partial \tilde S_{0u} \left( {t,r} \right)}}{{\partial r}} + \frac{{\partial \tilde S_{0u} \left( {t,r} \right)}}{{\partial t}}\frac{{dt}}{{dr}}\,.
\ee
In this section we don't use the Painleve transformation as we have used in the previous section.  Therefore, the corresponding radial null geodesic in the background (\ref{metric-gen-timedep}) is
\be \label{radial-null}
 \frac{{dr}}{{dt}} =  \pm \sqrt {XY}\,. 
\ee 
The $+(-)$ signs in the left hand side of (\ref{radial-null}) refers to the geodesic of outgoing(ingoing) null particles respectively. Combining equations (\ref{eqtn-Su-afterchain}) and (\ref{radial-null}) gives us
\be 
\frac{{\partial \tilde S_{0u} \left( {t,r} \right)}}{{\partial r}} = \frac{{d\tilde S_{0u} \left( {t,r} \right)}}{{dr}} \mp \frac{1}{\sqrt {XY}} \frac{{\partial \tilde S_{0u} \left( {t,r} \right)}}{{\partial t}}
\ee
with $ -(+)$ signs refer to the outgoing(ingoing) spin up Dirac from the black hole. 

Recall that for the dynamics of a particle with Hamiltonian $H$ and action $S$, one can show the relation \cite{Goldstein}
\be\label{S-H}
\frac{{\partial S}}{{\partial t}} + H = 0\,.
\ee
The last equation also emerges in the semiclassical discussion, for example, in WKB approximation to solve the one dimensional Schroedinger equation 
\be\label{Schro-wkb}
i\hbar \frac{{\partial \Psi }}{{\partial t}} =  - \frac{{\hbar ^2 }}{{2m}}\frac{{\partial ^2 \Psi }}{{\partial x^2 }} + V\left( x \right)\Psi \,,
\ee 
where we use the ansatz $\Psi  = e^{iS/\hbar } $ and $S$ is the classical action of particle associated to the wave function $\Psi$. From eq. (\ref{Schro-wkb}), we may observe that the partial derivative of action with respect to time would be a negative quantity for a particle with positive energy, since the eigenvalue of $H$ must be positive. Therefore, the $(-)$ sign in (\ref{eqfn}) belongs to the outgoing particle, 
\be\label{dtS0-minXYdrS0}
\partial _t S_{0u}  =  - \sqrt {XY} \partial _r S_{0u} \,,
\ee 
since the momentum $p_r = \partial_r S_0$ is positive. Correspondingly, the one with $(+)$ sign refers to the ingoing particle,
\be\label{dtS0-plusXYdrS0}
\partial _t S_{0u}  =  \sqrt {XY} \partial _r S_{0u} \,.
\ee 
Then we use equations (\ref{drS-dtS}), (\ref{eqtn-Su-afterchain}), (\ref{dtS0-minXYdrS0}), and (\ref{dtS0-plusXYdrS0}) to get
\be 
\frac{{d\tilde S_{0u} \left( {t,r} \right)}}{{dr}} =  \pm  {\frac{E\left( t \right)}{\sqrt{XY}}} 
\ee
whose solution can be read as
\be 
\tilde S_{0u} \left( {t,r} \right) =  \pm E\left( t \right)\int { {\frac{dr}{\sqrt{XY}}}} \,.
\ee
The $+$ and $-$ signs in the last equation belong to the outgoing and ingoing particle respectively.  Accordingly, a solution for the action (\ref{Su-gen}) can be read as
\be\label{S0usol-Dirac}
S_{0u} \left( {t,r} \right) =  - \int\limits_0^t {E\left( {t'} \right)dt'}  \pm \frac{{i\pi E\left( t \right)}}{{\sqrt {X'Y'} }}\,.
\ee 

Plugging the solution (\ref{S0usol-Dirac}) into (\ref{ansatzPsi-up}) gives us
\be
\Psi _{u,in}  = \left( {\begin{array}{*{20}c}
   {\cal{A}}  \\
   0  \\
   {\cal{B}}  \\
   0  \\
\end{array}} \right)\exp \left( {  \frac{i}{\hbar }\left( { - \int\limits_0^t {E\left( {t'} \right)dt'}  - \frac{{i\pi E\left( t \right)}}{{\sqrt {X'Y'} }}} \right)} \right)\,,
\ee 
and
\be 
\Psi _{u,out}  = \left( {\begin{array}{*{20}c}
   {\cal{A}}  \\
   0  \\
   {\cal{B}}  \\
   0  \\
\end{array}} \right)\exp \left( {  \frac{i}{\hbar }\left( { - \int\limits_0^t {E\left( {t'} \right)dt'}  + \frac{{i\pi E\left( t \right)}}{{\sqrt {X'Y'} }}} \right)} \right).
\ee 
Making the ingoing probability $P_{in} = \left| {\Psi _{u,in} } \right|^2 $ is unity, i.e. all fields that come close to a black hole will be absorbed, yields
\be
\int\limits_0^t {E\left( {t'} \right)dt'}  =  - \frac{{i\pi E\left( t \right)}}{{\sqrt {X'Y'} }}\,.
\ee 
Therefore the outgoing probability can be written as
\be 
P_{out}  = \left| {\Psi _{u,out} } \right|^2  = \exp \left( { - \frac{{4\pi E\left( t \right)}}{{\hbar \sqrt {X'Y'} }}} \right)\,.
\ee
The ``detailed balance'' principle tells us that 
\[P_{out}  = e^{ - \beta E} P_{in} \]
which then allow us to write the Hawking temperature for a general time dependent black hole (\ref{metric-gen-timedep}) as
\be\label{Hawking-temp-Dirac}
T_H  = \frac{{\hbar \sqrt {X'\left( {t,r_h } \right)Y'\left( {t,r_h } \right)} }}{{4\pi }}\,.
\ee 
The Hawking temperature (\ref{Hawking-temp-Dirac}) is interpreted as the measured temperature by a detector at infinity where the radiation consists of massless quantum particle with spin $\tfrac{1}{2}$ moving outward radially from the black holes.

\section{Photon Tunneling}\label{sec:photon}

In \cite{Majhispin}, Majhi and Samanta discuss the tunneling of photon and gravitino which yield the Hawking radiation from a static black hole. One of the conclusions in their work is that the Hawking radiation in the form of the tunneling of photon and gravitino yields the Hawking temperature which is invariant compared to the one computed in the case of scalar tunneling. In this section and the next one, we show that the same conclusion is obtained for a time dependent black hole. We start from an action for Maxwell fields in curved spacetime,
\be 
S =  - \frac{1}{4}\int {\sqrt { - g} F_{\mu \nu } F^{\mu \nu } } d^4 x\,.
\ee
Taking the variation of $A_\mu$ in the action above, we obtain
\be\label{Maxwelleqtn}
\nabla _\mu  F^{\mu \nu }  = 0\,,
\ee 
which is known as the Maxwell equation in the absence of the source $J^\nu$. By following Majhi et al \cite{Majhispin}, we use the Hamilton-Jacobi ansatz for the vector field
\be \label{Amu}
A^\mu   \sim k^\mu  e^{ \frac{i}{\hbar }S\left( {t,r,\theta ,\phi } \right)} \,,
\ee
where $k^\mu$ is the polarization vector. This polarization vector is independent of the spacetime coordinates. As usual, the action is expanded as
\be \label{S-expand}
S\left( {t,r,\theta ,\phi } \right) = \sum\limits_{i = 0}^\infty  {\hbar ^i S_i \left( {t,r,\theta ,\phi } \right)} 
\ee 
just like what we did in the spinor case (\ref{expandS-Dirac}). In \cite{Majhispin}, the polarization vector $k^\mu$ is also expanded in $\hbar$ since the authors discuss the quantum correction which comes from the higher order terms in $\hbar$ of $S(t,r,\theta,\phi)$ and $k^\mu$. However, since we are not interested in pursuing such quantum correction, the polarization vector $k^\mu$ in (\ref{Amu}) can be kept up to its semiclassical value only. 

Plugging the ansatz (\ref{Amu}) for the gauge fields into the equation (\ref{Maxwelleqtn}) which alternatively can be expressed as
\be \label{Maxwelleqtn2}
\partial _\mu  F^{\mu \nu }  + \Gamma _{\tau \mu }^\mu  F^{\tau \nu }  + \Gamma _{\tau \mu }^\nu  F^{\mu \tau } = 0\,,
\ee 
one can get
\be \label{eqtnnofix}
\left( {k^\nu  \partial ^\mu  S_0  - k^\mu  \partial ^\nu  S_0 } \right)\partial _\mu  S_0  = 0\,.
\ee
In getting the last equation, we have taken the limit $\hbar \to 0$ in the equation (\ref{Maxwelleqtn2}). We choose to work in the Lorentz gauge, 
\be \label{Lorentz}
\nabla _\mu  A^\mu   = 0\,,
\ee 
which after plugging the gauge fields (\ref{Amu}) gives
\be \label{Lorentz2}
k^\mu  \partial _\mu  S_0  = 0\,.
\ee 
Again we have employed the limit $\hbar \to 0$ in obtaining the equation (\ref{Lorentz2}). In this Lorentz gauge condition, the reading of equation (\ref{eqtnnofix}) reduces to 
\be \label{eqtnfixed}
k^\nu  \left( {\partial ^\mu  S_0 } \right)\left( {\partial _\mu  S_0 } \right) = 0\,.
\ee

Working on the $t$ and $r$ sectors only in the spacetime under consideration allow us to write (\ref{eqtnfixed}) as
\be 
g^{tt} \left( {\partial _t S_0 } \right)^2  + g^{rr} \left( {\partial _r S_0 } \right)^2  = 0 \,.
\ee 
We find that the last equation is similar to (\ref{eqfn}) if we replace $S_0$ with $S_{0u}$. It is clear since the blackground of spacetime where the vector probes come and fall into a black hole is also the same, i.e. the metric (\ref{metric-gen-timedep}). Therefore, the solution for $S(t,r,\theta,\phi)$ for the vector fields $A_\mu$ can be read as
\be
S_{0} \left( {t,r} \right) =  - \int\limits_0^t {E\left( {t'} \right)dt'}  \pm \frac{{i\pi E\left( t \right)}}{{\sqrt {X'Y'} }}\,.
\ee 

Accordingly, the ingoing and outgoing solutions for the vector fields can be read as
\be 
A_{{\rm{in}}}^\mu   \sim k^\mu  \exp \left( { \frac{i}{\hbar }\left( {-\int\limits_0^t {E\left( {t'} \right)dt'}  - i\pi \frac{{E\left( t \right)}}{{\sqrt {X'Y'} }}} \right)} \right)
\ee
and
\be 
A_{{\rm{out}}}^\mu   \sim k^\mu  \exp \left( {\frac{i}{\hbar }\left( {-\int\limits_0^t {E\left( {t'} \right)dt'}  + i\pi \frac{{E\left( t \right)}}{{\sqrt {X'Y'} }}} \right)} \right)
\ee
respectively. 
The unit incoming probability $P_{{\rm{in}}} = |A^\mu_{\rm{in}}|^2$, the relation between incoming and outgoing probabilities
\be 
P_{{\rm{out}}}  = P_{{\rm{in}}}  \exp \left( { - \frac{{4\pi E\left( t \right)}}{{\hbar \sqrt {X'Y'} }}} \right)\,,
\ee
and the ``detailed balance'' principle $P_{{\rm{out}}} = P_{{\rm{in}}} \exp{(-\beta E(t))}$ yields the reading of Hawking temperature is
\be 
T_H  = \frac{{\hbar \sqrt {X'\left( {t,r_h } \right)Y'\left( {t,r_h } \right)} }}{{4\pi }}\,.
\ee
One observes that the temperature in the last equation is equal to the one computed in the Dirac particle case (\ref{Hawking-temp-Dirac}) and scalar case \cite{Siahaan:2009qv}.

\section{Gravitino tunneling}\label{sec:gravitino}

We start with the action of massless Rarita-Schwinger $\Psi_\alpha$ fields in curved spacetime,
\be\label{RS-action} 
S_\psi   = \int {d^4 x\sqrt { - g} \bar \Psi^\alpha i\tilde \gamma ^\mu  \left( {\partial _\mu   + \frac{i}{2}g^{\gamma \nu } \Gamma _{\mu \nu }^\beta  \Sigma _{\beta \gamma } } \right)\Psi_\alpha } \,,
\ee
where $\Sigma_{\beta\gamma}$ and the Dirac matrices $\tilde{\gamma}^\mu$ in the action above are those used in the Dirac action (\ref{Dirac-action}). 
Accordingly, the action (\ref{RS-action}) tells us that the corresponding equation of motion for $\Psi_\alpha$ can be read as
\be \label{RSeqtn}
{\tilde{\gamma}} ^\mu  \nabla _\mu  \Psi_\alpha  = 0\,,
\ee
which is known as the massless Rarita-Schwinger equation in curved space. It looks like the Dirac equation, with the Dirac spinor $\Psi$ is replaced by the vector-spinor $\Psi_\mu$. The number of degree of freedom of $\Psi_\mu$ is sixteen, which eight of them are removed by the two additional constraints: $\tilde{\gamma}^\mu\Psi_\mu = 0$ and $\nabla^\mu\Psi_\mu = 0$. 

The Hamilton-jacobi ansatz for vector-spinor $\Psi_\mu$ can be read as
\be 
\Psi _{\left( u \right)\mu }  = \left( {\begin{array}{*{20}c}
   {{\cal{A}}_\mu  }  \\
   0  \\
   {{\cal{B}}_\mu  }  \\
   0  \\
\end{array}} \right)\exp \left( { \frac{i}{\hbar }S_{\left( u \right)} } \right)\,,
\ee
and
\be 
\Psi _{\left( d \right)\mu }  = \left( {\begin{array}{*{20}c}
   0  \\
   {{\cal{C}}_\mu  }  \\
   0  \\
   {{\cal{D}}_\mu  }  \\
\end{array}} \right)\exp \left( { \frac{i}{\hbar }S_{\left( d \right)} } \right)\,,
\ee
where $\Psi _{\left( u \right)\mu }$ and $\Psi _{\left( d \right)\mu }$ are the Rarita-Schwinger fields with spins $+3/2$ and $-3/2$ respectively. In the background (\ref{metric-gen-timedep}), equation (\ref{RSeqtn}) for radial geodesic can be read as
\be\label{dRS1}
\frac{{i{\cal{A}}_\mu}}{{\sqrt X }}\partial _t S_{0(u)}  + {\sqrt Y }{\cal{B}}_\mu\partial _r S_{0(u)}  + {\cal{O}}\left( \hbar  \right) = 0\,,
\ee
\be\label{dRS3}
{\sqrt Y }{\cal{A}}_\mu\partial _r S_{0(u)}  - \frac{{i{\cal{B}}_\mu}}{{\sqrt X }}\partial _t S_{0(u)}  + {\cal{O}}\left( \hbar  \right) = 0\,.
\ee
The action $S_{0(u)}$ is understood as the zeroth order term in the action expansion $S_{\left( u \right)}  = \sum\limits_{i = 0}^\infty  {\hbar ^i S_{i\left( u \right)} } $. Moreover, the last two equations are very close to (\ref{d1}) and (\ref{d3}) since the operator that applies to the massless Rarita-Schwinger fields in (\ref{RSeqtn}) is just the same with that applies to the massless Dirac field in (\ref{Diraceqtn}). Analogous to the technique applies to the Dirac fermion in the previous section, we rewrite the equations (\ref{dRS1}) and (\ref{dRS3}) in the form
\[
\left( {\begin{array}{*{20}c}
   {iX^{ - 1/2} \partial _t S_{0(u)} } & {Y^{ - 1/2} \partial _r S_{0(u)} }  \\
   {Y^{ - 1/2} \partial _r S_{0(u)} } & { - iX^{ - 1/2} \partial _t S_{0(u)} }  \\
\end{array}} \right)\left( {\begin{array}{*{20}c}
   {\cal{A}}_\mu  \\
   {\cal{B}}_\mu  \\
\end{array}} \right) \equiv {\tilde D}\left( {\begin{array}{*{20}c}
   {\cal{A}}_\mu  \\
   {\cal{B}}_\mu  \\
\end{array}} \right) = 0\,,
\]
whose solution for the action $S_{0(u)}$ finally can be found as
\be\label{S0usol-RS}
S_{0(u)} \left( {t,r} \right) =  - \int\limits_0^t {E\left( {t'} \right)dt'}  \pm \frac{{i\pi E\left( t \right)}}{{\sqrt {X'Y'} }}\,.
\ee 
Procedure that is needed in obtaining the solution (\ref{S0usol-RS}) is obvious since we deal the same equations in section \ref{sec:Dirac}. In discussing gravitino, we replace the complex valued functions ${\cal{A}}$ and ${\cal{B}}$ with the vectors ${\cal{A}}_\mu$ and ${\cal{B}}_\mu$ which are also complex valued. Moreover, the solutions for spin $\tfrac{3}{2}$ fields then can be written as
\be
\Psi _{\mu (u,in)}  = \left( {\begin{array}{*{20}c}
   {\cal{A}}_\mu  \\
   0  \\
   {\cal{B}}_\mu  \\
   0  \\
\end{array}} \right)\exp \left( { \frac{i}{\hbar }\left( { - \int\limits_0^t {E\left( {t'} \right)dt'}  - \frac{{i\pi E\left( t \right)}}{{\sqrt {X'Y'} }}} \right)} \right)\,,
\ee 
and
\be 
\Psi _{\mu (u,out)}  = \left( {\begin{array}{*{20}c}
   {\cal{A}}_\mu  \\
   0  \\
   {\cal{B}}_\mu  \\
   0  \\
\end{array}} \right)\exp \left( { \frac{i}{\hbar }\left( { - \int\limits_0^t {E\left( {t'} \right)dt'}  + \frac{{i\pi E\left( t \right)}}{{\sqrt {X'Y'} }}} \right)} \right)\,.
\ee 
Again, using the ''detailed balance'' principle for the relation of outgoing and ingoing gravitino probabilities one can get 
\be
T_H  = \frac{{\hbar \sqrt {X'\left( {t,r_h } \right)Y'\left( {t,r_h } \right)} }}{{4\pi }}\,,
\ee 
as the Hawking temperature from dynamical black holes (\ref{metric-gen-timedep}) in the form of massless gravitino tunneling from black holes. We observe that the Hawking temperature due to the massless gravitino tunneling is equal to the temperatures computed in the last two sections.

\section{Conclusion and Discussions}\label{sec:conclusion}

We have analyzed the Hawking radiation in the form of Dirac fermion, photon, and gravitino tunneling across the event horizon of time dependent black holes. The resulting Hawking temperatures are invariant compared to the one obtained in the scalar tunneling case. The results lead to a conclusion that the Hawking temperature obtained in the tunneling method is independent of the spins of the tunneled particles. We confirm that for a time dependent and spherically symmetric black hole whose metric has the form (\ref{metric-gen-timedep}), the Hawking temperature is independent of the spins of the tunneled particle, as in the case of static one.

It is interesting to note that the PW method presented in section \ref{sec:radial-null} works for a limited case only, i.e. the spacetime metric whose metric functions satisfy the condition (\ref{int.cond}). The obtained Hawking temperature in this method is confirmed by the result derived via Hamilton-Jacobi method by using higher spins tunneling in the succeeding sections as well as scalar tunneling \cite{Siahaan:2009qv}. In the Hamilton-Jacobi method we do not make use of the Painleve transformation, hence there is no integrability condition (\ref{int.cond}) to be fulfilled. Such condition also does not appear in \cite{Nielsen:2008kd}, and yet the same result for Hawking temperature (\ref{Hawking-temp}) is achieved by the authors. Hence, presumably there are some other transformations which yield the line element (\ref{metric-gen-timedep}) to be regular at event horizon, and therefore the PW method can be performed. Clearly each of these transformations will have a set of integrability conditions which justify a class of spacetime which fits in the computation of Hawking temperature using PW method. Finding an alternative to the generalized Painleve transformation (\ref{dtpdt}) and (\ref{dtpdr}) which yields the metric (\ref{metric-gen-timedep}) transforms to be regular at the event horizon would be challenging and we address this issue in our future work.

In their seminal paper \cite{Wilczek}, Parikh and Wilczek included the back reaction effect in their analysis of Hawking radiation in the tunneling picture. Hence, discussing a correction to the entropy of time dependent black holes which comes from the back reaction effect should also be possible. For the case of Vaidya black holes entropy, Zhang et al \cite{Zhang:2007ar} had carried out an analysis which takes back reaction effect into account, where the Vaidya spacetime is written in the Eddington-Finkelstein coordinate rather that Schwarzschild-like one. We note that Zhang et al consider the particle energy to be time independent, unlike the time dependent case presented in this paper. We address the analysis of time dependent black hole's entropy by considering back reaction effect in Schwarzschild-like coordinate in our future project.

\section*{Acknowledgments}
I thank Profs. Paulus Tjiang and Triyanta for useful discussions.

\appendix
\section{Example: Vaidya black holes}\label{app.Vaidya}

The Vaidya black hole is an interesting object to be studied \cite{Vaidya-rel}. It is an example of exact time dependent black hole solution where it solves the Einstein equations with $T_{\mu \nu }  = \rho k_\mu  k_\nu $. Since $k_\mu  k^\mu   = 0$, one can consider that the Vaidya solution describes a non empty spacetime outside of a null radiating mass. Scalar tunneling as a mechanism for Hawking radiation from a Vaidya black hole is also discussed in \cite{Ren:2006zu}, and the authors of \cite{Li:2008ap} study the spin $\tfrac{1}{2}$ tunneling as a Hawking process for this black hole.

Related to the generalized Painleve transformation (\ref{dtpdt}) and (\ref{dtpdr}), we can take Vaidya black holes \cite{Vaidya,Siahaan:2009qv},
\[ds^2  =  - \left( {\frac{{\dot m\left( {t,r} \right)}}{{f\left( {t,r} \right)}}} \right)^2 \left( {1 - \frac{{2m\left( {t,r} \right)}}{r}} \right)dt^2 \]
   \be
    + \left( {1 - \frac{{2m\left( {t,r} \right)}}{r}} \right)^{ - 1} dr^2  + r^2 d\Omega _2^2 \,.
   \ee 
   as a concrete example. In the last equation, $f(t,r)$ is an arbitrary function that depends on the model of mass $m(t,r)$ being used and satisfies
   \be
   f\left( {t,r} \right) = \left( {1 - \frac{{2m\left( {t,r} \right)}}{r}} \right)\frac{{\partial m\left( {t,r} \right)}}{{\partial r}}\,.
   \ee 
   It follows that from eq. (\ref{int.cond}), the mass functions of Vaidya black hole which are compatible with the time dependent Painleve transformation (\ref{dtpdt}) and (\ref{dtpdr}) should obey an equation which reads
   \be\label{condt.vaidya}
   2m\left( {t,r} \right)f\left( {t,r} \right)^2  = r\dot m\left( {t,r} \right)^2 \left( {1 - \frac{{2m\left( {t,r} \right)}}{r}} \right)^2 C\left( r \right)\,.
   \ee 
   Hence, the case of Vaidya black holes whose Hawking temperature can be computed using the method proposed in this paper are those with the mass functions that satisfy eq. (\ref{condt.vaidya}), which is found to be
\[
T_H  = \left\{ {\frac{{\hbar \left( {m - rm'} \right)}}{{2\pi r^2 }}\left[ {\left( {\frac{{\dot m}}{f}} \right)^2 } \right.} \right.
\]
\be 
\left. {\left. { + \left( {\frac{{\dot m}}{f}} \right)\frac{{\partial \left( {\dot m/f} \right)}}{{\partial r}}\frac{{r\left( {r - 2m} \right)}}{{m - rm'}}} \right]^{1/2} } \right\}_{r = r_h \left( t \right)} \,.
\ee
Note that at the Schwarzschild limit \cite{Vaidya}, where $f(t,r) = - {\dot m}$, the Hawking temperature above reduces to the familiar $T_H = \frac{\hbar}{8\pi m}$ for Schwarzschild black hole.

\section{Time dependent Hamiltonian in Schrodinger equation}\label{app.timedepSchrodinger}
In this appendix we elaborate some subtleties in time dependent Schrodinger equation in supporting the general form of solution for action (\ref{Su-gen}). The analysis presented here can be found in \cite{Dittrich} chapter 31. 

Let us consider a Hamiltonian $H(r(t))$ which describes a physical system. The variable $r(t)$ are parameters which slowly vary with respect to time $t$. Hence, we understand that the Hamiltonian under consideration here is time dependent. Accordingly, the Schrodinger equation to be solved in this case is
\be \label{schro.app}
i\hbar \frac{\partial }{{\partial t}}\left| {\Psi \left( t \right)} \right\rangle  = H\left( {r\left( t \right)} \right)\left| {\Psi \left( t \right)} \right\rangle \,.
\ee
The associated Hamiltonian operator and energy eigenvalues relation can be read as
\be 
H\left( {r\left( t \right)} \right)\left| {k,r\left( t \right)} \right\rangle  = E_k \left( {r\left( t \right)} \right)\left| {k,r\left( t \right)} \right\rangle 
\ee
which is valid at an instantaneous time $t$ only. The state $\left| {k,r\left( t \right)} \right\rangle $ is normalized accordingly to
\be
\left\langle {{k,r\left( t \right)}}
 \mathrel{\left | {\vphantom {{k,r\left( t \right)} {l,r\left( t \right)}}}
 \right. \kern-\nulldelimiterspace}
 {{l,r\left( t \right)}} \right\rangle  = \delta _{kl} 
\ee 

The solution to equation (\ref{schro.app}) can be expressed as
\be\label{psi.app}
\left| {\Psi \left( t \right)} \right\rangle  = \sum\limits_k {a_k \left( t \right)e^{ - \frac{i}{\hbar }\int\limits_0^t {dt' E_k \left( {r\left( t'  \right)} \right)} } } \left| {k,r\left( t \right)} \right\rangle \,.
\ee 
Plugging (\ref{psi.app}) into the Schrodinger equation (\ref{schro.app}) gives us
\be
\sum\limits_k {e^{ - \frac{i}{\hbar }\int\limits_0^t {dt' E_k \left( {r\left( t'  \right)} \right)} } \left\{ {\dot a_k  + a_k \frac{\partial }{{\partial t}}} \right\}} \left| {k,r\left( t \right)} \right\rangle  = 0\,.
\ee 
Using the adiabatic approximation, the solution to $a_k (t)$ can be read as\footnote{A detail derivation can be found in \cite{Dittrich}.} 
\be 
a_k \left( t \right) = e^{iG_k \left( t \right)} \,,
\ee
where $G_k\left( t \right)$ is some purely real time dependent function, i.e.
\be
G_k \left( t \right) = i\int\limits_0^t {dt' \left\langle {k,r\left( t'  \right)} \right|} \frac{\partial }{{\partial t' }}\left| {k,r\left( t'  \right)} \right\rangle \,.
\ee 
Finally, the solution to Schrodinger equation (\ref{schro.app}) in the adiabatic approximation can be read as
\be
\left| {\Psi \left( t \right)} \right\rangle  = e^{\frac{i}{\hbar }\int\limits_0^t {dt' \left\{ { - E_n \left( {r\left( t'  \right)} \right) + \hbar \left\langle {k,r\left( t'  \right)} \right|\frac{\partial }{{\partial t' }}\left| {k,r\left( t'  \right)} \right\rangle } \right\}} } \left| {k,r\left( t  \right)} \right\rangle \,,
\ee 
which justifies the general form of action's solution (\ref{Su-gen}).
%
%

\end{document}